# Excitation Mechanisms of Er Optical Centers in GaN Epilayers

D. K. George,[1] M. Hawkins,[1] M. McLaren,[1] H. X. Jiang,[2] J. Y. Lin,[2] J. M. Zavada,[3] and N. Q. Vinh[1,a]

[1]Department of Physics, Virginia Tech, Virginia 24061, USA

[2]Department of Electrical and Computer Engineering, Texas Tech University, Lubbock, Texas 79409, USA

[3]Department of Electrical and Computer Engineering, NYU Polytechnic School of Engineering, Brooklyn, New York 11201, USA

Abstract

We report direct evidence of two mechanisms responsible for the excitation of optically active $Er^{3+}$ ions in GaN epilayers grown by metal-organic chemical vapor deposition. These mechanisms, resonant excitation via the higher-lying inner 4*f* shell transitions and band-to-band excitation of the semiconductor host, lead to narrow emission lines from isolated and the defect-related Er centers. However, these centers have different photoluminescence spectra, decay dynamics, and excitation cross sections. The isolated Er optical center, which can be excited by either mechanism, has the same decay dynamics, but possesses a much higher cross-section under band-to-band excitation. In contrast, the defect-related Er center can only be excited through band-to-band excitation but has the largest cross-section. These results explain the difficulty in achieving gain in Er doped GaN and indicate new approaches for realization of optical amplification, and possibly lasing, at room temperature.

a) Electronic email: Vinh@vt.edu

Rare earth (RE) doped semiconductors have played an important role in various optoelectronic and photonic applications, ranging from emitting elements in solid-state lasers and in phosphors for color lamps and displays[1-3] to optical fiber telecommunications[4] and to quantum information processing[5-7]. Among the various rare earth elements, Er has attracted particular attention because the $^4I_{13/2} \rightarrow {^4I_{15/2}}$ transition involving nonbonding 4$f$ shell electrons of the Er$^{3+}$ ion occurs at the technologically important wavelength of 1.54 µm, matching the absorption minimum of silica-based optical fibers[1-4]. GaN is expected to be an ideal host material for RE doping because it is a wide and direct bandgap semiconductor, which exhibits less thermal quenching and stronger RE emission at room temperature than RE-doped Si[8-11]. As a result of a continuing research effort, GaN:Er-based light emitting diodes have now been successfully developed[12, 13]. In spite of impressive developments in this area, the GaN:Er system remains poorly understood and even controversial in regard to the excitation mechanisms underlying the luminescent of Er$^{3+}$ ions which constitutes a barrier to further increases of device emission efficiency and thermal stability.

The excitation of Er$^{3+}$ ions can take several steps before reaching the luminescence of the ions. A common way of exciting Er$^{3+}$ which has been done in most RE-based lasers is to employ an optical pumping source with a photon energy that matches a higher-lying inner 4$f$ shell transition. This type of excitation scheme is therefore called direct or resonant excitation. The method needs a high optical pumping power due to a low excitation cross-section for RE ions in insulators with a typical value of $10^{-20}$ cm$^2$.[4] Other types of excitation mechanisms involve host excitation first and then Er$^{3+}$ ions using a band-to-band excitation. The common feature to these excitation processes involves the recombination of electrons and holes with a non-radiative energy transfer to nearby Er$^{3+}$ ions. These processes are called indirect or non-resonant band-to-band excitation with an efficiency about three to five orders of magnitudes higher than the resonant process. Both electron and hole carriers can be bound to each other forming either a free exciton or a bound exciton (BE) trapped by RE ions or an impurity/local defect nearby RE ions. In this paper, we report direct evidence of two different mechanisms for two Er optical centers in GaN:Er epilayers prepared by metal-organic chemical vapor deposition (MOCVD) using direct and indirect excitation processes.

The Er doped GaN epilayers in this study were synthesized by MOCVD in a horizontal reactor on (0001) $c$-plane sapphire substrates[14]. Two different types of Er-doped GaN structures were realized: a GaN:Er epilayer of 0.5 µm thickness on a thin un-doped GaN layer of 1.2 µm, and a GaN:Er epilayer of 0.8 µm thickness on a thin un-doped GaN layer of 200 nm on a top of an AlN layer of 0.5 µm thickness. The Er concentration ($n_{Er}$) was $2 \times 10^{20}$ cm$^{-3}$. Following MOCVD growth, the samples were characterized by X-ray diffraction and photoluminescence (PL) measurements. The powder XRD spectra indicated high crystallinity and no second phase formation.

The PL spectra were obtained with a high resolution spectroscopy (Horiba iHR550) equipped with a 900 grooves/mm grating blazed at 1.5 µm and detected by a high sensitivity liquid nitrogen InGaAs detector (DSS-IGA). In this specific case the PL spectrum resolution is 0.1 nm. Optical measurements were performed using a closed-cycle optical cryostat (Janis) accessing the 10 – 300 K range. The time-resolved measurements were carried out using a Techtronix TDS 3000 digital oscilloscope in combination with the InGaAs detector. The time resolution was measured as 50 µs.

The high resolution PL spectra under resonant and non-resonant band-to-band excitation of Er optical centers in GaN allow us to determine the site symmetry of these centers. For the resonant excitation, Er ions are excited resonantly at $\lambda_{exc}$ = 809 nm within the $Er^{3+}$ $^4I_{15/2} \rightarrow {}^4I_{9/2}$ transition using a Ti:Sapphire laser. The spectrum of Er optical ions at 15 K composed of a few sharp PL lines at 1.54 µm corresponds to transitions between Stark sublevels from $^4I_{13/2} \rightarrow {}^4I_{15/2}$ intra-4$f$ shell transitions (Fig. 1). We also have used a non-resonant wavelength of 820 nm from the Ti:Sapphire laser to excite the Er optical centers, but the PL signal is very weak under the detection level. Unlike a previous report on the Er in GaN with an implantation method,[15] our PL spectrum contains a few narrow PL lines originated from a single Er optical center in GaN. Excitation of Er optical centers with photon energy higher than the GaN bandgap for the non-resonance band-to-band excitation using an Argon laser operating at 351.1 and 363.8 nm leads to a different PL spectrum (Fig. 1). The difference between resonant and non-resonant band-to-band spectra allows us to identify two types of optical centers: (1) Er ions in an isolated local environment can be excited via the resonant excitation $^4I_{15/2} \rightarrow {}^4I_{9/2}$ transition and also the non-resonant excitation due to the recombination of electrons and holes with a non-radiative energy transfer. This optical center is referred to an isolated Er optical center labeled L with PL lines $L_1$, $L_{2A}$, $L_{2B}$, $L_3$. (2) Er ions strongly associated with nearby defects or impurities can be excited indirectly via the host involving a trapped (bound) exciton. This center is referred to a defect-related Er optical center labeled L' with PL lines $L'_{1A}$, $L'_{1B}$, $L'_2$, $L'_3$, $L'_4$, $L'_5$. The spectrum under band-to-band excitation has more narrow PL lines. The PL lines come from all Er optical centers (isolated and defect-related Er optical centers) in GaN. The difference of PL spectra indicates a difference of the local site symmetry for two Er optical centers under different excitation processes. Line positions can be found in Table 1.

An ab-initio calculation[16] and Rutherford Backscattering Spectroscopy experiment study[17] have shown that the majority of Er ions in GaN mostly occupy Ga substitutional sites. Typically, RE ions in the substitutional sites were considered as an isoelectronic impurity center.[18] The isoelectronic center has no net charge in the local bonding region. A hole or an electron can be localized at such a center by a local core potential; subsequently, the secondary particle can be captured by Coulomb field of the first particle. The recombination of the two particles will transfer their energy to the Er ion. As can be seen from the Fig. 1, the isolated Er optical center can be excited by two mechanisms of resonant excitation via the $^4I_{15/2} \rightarrow$

$^4I_{9/2}$ transition and the non-resonant band-to-band excitation through the recombination of electron and hole pairs.

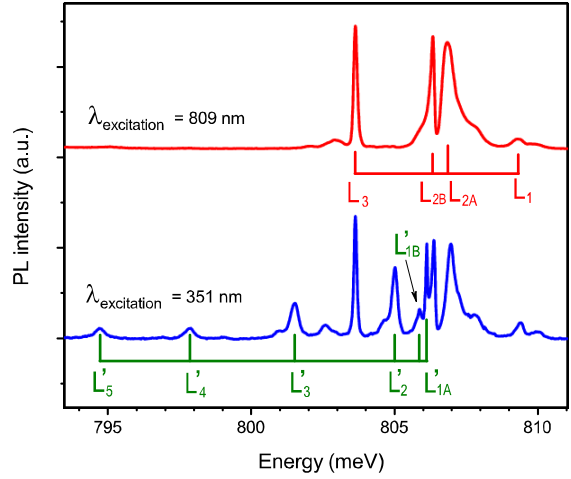

Figure 1: PL spectra of GaN:Er sample as measured at 15 K under resonant ($\lambda_{exc}$ = 809 nm) and non-resonant ($\lambda_{exc}$ = 351.1 nm) excitation. Line positions, decay times, and excitation cross sections can be found in Table 1.

Table I: Labeling, spectroscopic parameters, decay dynamics and excitation cross-section values of the Er optical centers in GaN sample at 10 K.

| Transition label | Energy (meV) | Decay time (ms) | Excitation cross-section (cm²) | |
|---|---|---|---|---|
| | | 351 nm | 351 nm | 809 nm |
| Isolated Er optical center | | | | |
| $L_1$ | 809.31 | 3.2 ± 0.3 | 9.1 × 10⁻¹⁷ | 3.1 × 10⁻²⁰ |
| $L_{2A}$ | 806.85 | 3.0 ± 0.3 | 9.5 × 10⁻¹⁷ | 3.0 × 10⁻²⁰ |
| $L_{2B}$ | 806.32 | 3.0 ± 0.3 | 9.8 × 10⁻¹⁷ | 3.0 × 10⁻²⁰ |
| $L_3$ | 803.62 | 3.1 ± 0.3 | 9.6 × 10⁻¹⁷ | 3.1 × 10⁻²⁰ |
| Defect-related Er optical center | | | | |
| $L'_{1A}$ | 806.11 | - | 4.1 × 10⁻¹⁶ | NA |
| $L'_{1B}$ | 805.85 | | | NA |
| $L'_2$ | 805.00 | 0.8 ± 0.3  2.1 ± 0.3 | 3.8 × 10⁻¹⁶ | NA |
| $L'_3$ | 801.51 | 0.7 ± 0.3  2.1 ± 0.3 | 3.5 × 10⁻¹⁶ | NA |
| $L'_4$ | 797.85 | - | - | NA |
| $L'_5$ | 794.72 | - | - | NA |

The PL for numerous RE ions in GaN has been observed under above or below bandgap excitation.[15, 19, 20] The excitation of Er centers is commonly described as involving (trapped) BEs with subsequent non-radiative transfer of the BE energy to nearby Er ions. The exact nature of the carrier trap is still under discussion, but is believed to be related to defects, impurities or defect–impurity complexes, rather than to RE ions acting as traps. The efficiency of this process is high, but the requirement of BEs for excitation will open non-radiative recombination channels for the luminescence process. A typical non-radiative recombination channel is the Auger de-excitation process which shows a fast lifetime on the decay dynamics.[21-23] With non-recombination channels, the PL intensity is quenched strongly with the increasing of temperature. Thus we cannot obtain the optical amplification in these optical centers at room temperature.[22]

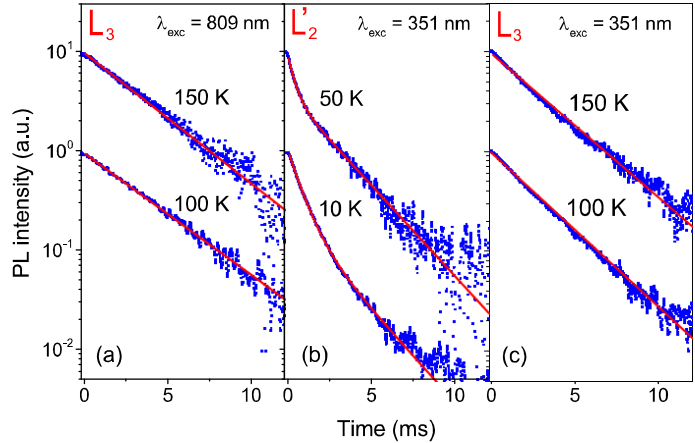

Figure 2: Decay dynamics of PL lines at different temperatures and excitation energies. (a) Under resonant excitation ($\lambda_{exc}$ = 809 nm), the decay dynamics of $L_3$ are temperature independent with a time of 3.3 ms at 100 and 150 K. (b) Under band-to-band excitation ($\lambda_{exc}$ = 351.1 nm), the dynamics for the line $L'_2$ belonging to the defect-related center shows a double-exponential function with 0.8 and 2.1 ms decay times at 10 and 50 K. (c) The dynamics of the line $L_3$ is independence with temperature of 3.1 ms, shown at 100 and 150 K.

In order to understand excitation mechanisms of Er optical centers, we have examined decay characteristics of the two optical centers under resonant as well as the non-resonant excitation (Fig. 2). Under resonant excitation, the dynamics of all PL lines of the isolated Er optical centers appear as a single experiential decay dynamics of 3.3 ± 0.3 ms (Fig. 2a). The value is similar with previous reports on the Er in GaN.[15, 24, 25] The decay dynamics we observe here is effectively independent of temperature over the entire range from 10 K to room temperature. It indicates that the isolated Er optical center has no quenching channels. Under band-to-band excitation, while again the isolated Er optical center reveals a similar decay time of 3.1 ± 0.3 ms and an independence with temperature as for the resonant excitation method (Fig. 2c), the defect-related Er optical center shows a double-exponential decay dynamics of 0.8 and 2.1 ms and strongly depends on temperature (Fig. 2b). Above 150 K, we cannot detect any PL lines for this optical center. The double-exponential decay process of the optical center indicates that there are non-radiative recombination processes taking place in the PL signal.

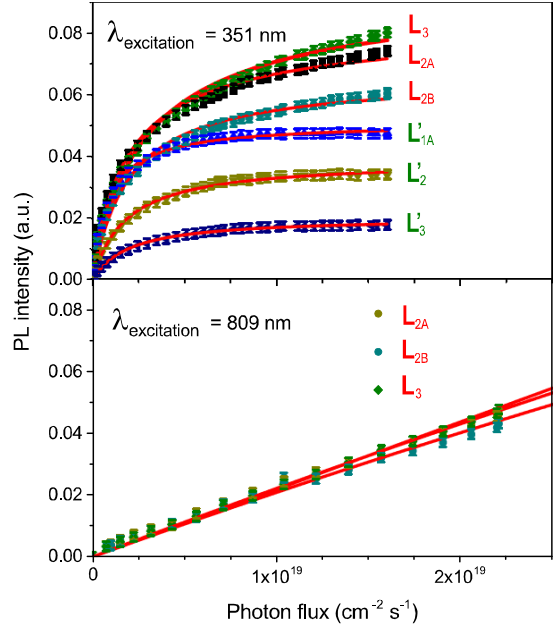

Figure 3: PL intensities measured at 10 K as a function of excitation flux for: (a) lines $L_{2A}$, $L_{2B}$ and $L_3$, from the isolated Er optical centers, and $L'_{1A}$, $L'_2$, $L'_3$, from the defect-related Er optical centers, under band-to-band excitation ($\lambda_{exc}$ = 351.1 nm); and, (b) lines $L_{2A}$, $L_{2B}$ and $L_3$ under resonant excitation ($\lambda_{exc}$ = 809 nm).

The Er PL in GaN under the band-to-band excitation is explained by the non-radiative recombination of electron-hole pairs (excitons) or bound excitons to the isolated or defect-related Er ions and the effective excitation cross-section is extensively used to qualify the excitation processes. The behavior of the excitation flux dependence of the PL intensity under cw laser for band-to-band excitation at 10 K is different for the two optical centers (Fig. 3a). Specifically, while the Er PL intensity from defect-related optical centers appears to saturate at lower photon flux, the Er PL intensity from the isolated optical centers obviously increases with photon flux. As has been discussed previously,[22, 23] under steady state conditions, the photon flux dependence of Er PL intensity is well described with the formula:

$$I_{\mathrm{PL}} \propto N_{Er}^* = \frac{N_{\mathrm{Er}}\sigma\tau\Phi}{1+\sigma\tau\Phi} \qquad (1)$$

where $N_{Er}^*$ is the concentration of optically active erbium ions, $\sigma$ is an effective excitation cross-section of $Er^{3+}$ ion, $\tau$ is the effective lifetime of $Er^{3+}$ in the excited state, $^4I_{13/2}$, $\Phi$ is the excitation photon flux, and $N_{Er}$ is the total concentration of erbium ions. The solid curves in Fig. 3(a) represent the best fits to the experimental data using Eq. 1. From the fitting, we have derived values of the effective excitation cross-section for the defect-related Er optical center, $\sigma'(\lambda_{exc}$ = 351.1 nm), equal to $(3.8 \pm 2) \times 10^{-16}$ cm$^2$ for PL lines $L'_{1A}$, $L'_2$, $L'_3$. This value is in agreement with the previous reports on the centers.[15] We also have extracted values of the effective excitation cross-section of the isolated Er optical center, $\sigma$, ($\lambda_{exc}$ = 351.1 nm), equal to $(9.5 \pm 2) \times 10^{-17}$ cm$^2$ for PL lines $L_{2A}$, $L_{2B}$, $L_3$. The isolated optical centers do not require a defect or an impurities center, thus the capture as well as recombination processes of electrons and holes are less efficient, and as a result $\sigma' > \sigma$. In a similar way, we also achieved a value of effective excitation cross-section for the resonant excitation $\sigma$ ($\lambda_{exc}$ = 809 nm), equal to $(3.0 \pm 2) \times 10^{-20}$ cm$^2$ (see Fig. 3b). This is a typical value for the intra-4$f$ shell excitation cross-section of RE in insulators[4] and many orders of magnitudes lower than those of the band-to-band excitation process.

We note that although the GaN:Er system exhibits a long excited state lifetime, due to the forbidden character of the intra-4$f$-electron shell transitions, population inversion and laser action have not been achieved in crystalline GaN:Er (while optical amplifiers based on Er-doped insulators are routinely manufactured). Realization of lasing action would provide a major boost for optoelectronic applications of GaN. In order to achieve gain, the absorption by Er$^{3+}$ ions should be maximized and losses minimized. The latter include absorption of the 1.5 µm radiation in the host and non-radiative recombinations of excited Er$^{3+}$ ions. The temperature independence of the long decay time of 3.0 ms and a similar dynamics with the resonant excitation process indicate that there is no non-radiative recombination process in the isolate Er optical center prepared by the MOCVD. In addition, by integrating all PL intensities from the isolated Er optical center in the PL spectrum under band-to-band excitation and taking into account the decay dynamics for both optical centers, we find that there is more than 70% contribution of the isolated Er optical centers to the PL spectrum.

In summary, we have investigated the excitation mechanisms of Er optical centers in GaN epilayer prepared by MOCVD. The PL spectra under resonant and non-resonant band-to-band excitation reveal an existence of two types of optical centers; the defect-related and the isolated Er optical centers. The defect-related optical centers appear a strong temperature dependence of PL intensity with a double-exponential decay dynamics and the highest excitation cross-section. This Er optical center can only be excited by the band-to-band excitation. The isolated Er optical centers which can be excited by both resonant and non-resonant processes show a temperature independence of decay dynamics ~3.0 ms. The excitation cross-section of this optical center under band-to-band excitation is much larger than that using the resonant excitation process via the $^4I_{15/2} \rightarrow {}^4I_{9/2}$ transition. The high excitation cross-section combined with the temperature independence of decay dynamics and high percentage of optically active center of the isolated Er optical centers makes the GaN:Er materials synthesized by MOCVD interesting for GaN photonics. In particular, the material appears promising for realization of population inversion and, consequently, optical amplification at room temperature.

NQV acknowledges support by NSF (ECCS-1358564). The materials growth effort at TTU was supported by JTO/ARO (W911NF-12-1-0330) and the related characterization work was partially supported by NSF (ECCS-1200168).